\documentclass[aps,prl,twocolumn,superscriptaddress,showpacs,longbibliography]{revtex4-1}
\usepackage[colorlinks=true,citecolor=blue,linkcolor=blue,breaklinks=true]{hyperref}
\usepackage{amssymb}
\usepackage{graphicx}
\usepackage{amsmath}
\usepackage{mathtools}
\usepackage[export]{adjustbox}
\usepackage{epsfig}
\usepackage{times}
\usepackage{color}
\usepackage{subfigure}
\usepackage{setspace}
\usepackage{bm}
\usepackage{calc}
\usepackage{natbib}

\begin{document}

\newcommand {\ba} {\ensuremath{b^\dagger}}
\newcommand {\Ma} {\ensuremath{M^\dagger}}
\newcommand {\psia} {\ensuremath{\psi^\dagger}}
\newcommand {\psita} {\ensuremath{\tilde{\psi}^\dagger}}
\newcommand{\lp} {\ensuremath{{\lambda '}}}
\newcommand{\A} {\ensuremath{{\bf A}}}
\newcommand{\Q} {\ensuremath{{\bf Q}}}
\newcommand{\kk} {\ensuremath{{\bf k}}}
\newcommand{\qq} {\ensuremath{{\bf q}}}
\newcommand{\kp} {\ensuremath{{\bf k'}}}
\newcommand{\rr} {\ensuremath{{\bf r}}}
\newcommand{\rp} {\ensuremath{{\bf r'}}}
\newcommand {\ep} {\ensuremath{\epsilon}}
\newcommand{\nbr} {\ensuremath{\langle ij \rangle}}
\newcommand {\no} {\nonumber}
\newcommand{\up} {\ensuremath{\uparrow}}
\newcommand{\dn} {\ensuremath{\downarrow}}
\newcommand{\rcol} {\textcolor{red}}

\begin{abstract}

  We propose a new field theoretic method for calculating Renyi entropy of a sub-system of many interacting Bosons without using replica methods. This method is applicable to dynamics of both open and closed quantum systems starting from arbitrary initial conditions.
 Our method identifies the Wigner characteristic of a reduced density matrix with the partition function of the whole system with a set of linear sources turned on only in the subsystem and uses this to calculate the subsystem's Renyi entropy.
  We use this method to study evolution of Renyi entropy in a non-interacting open quantum system starting from an initial Fock state. We find a relation between the initial state and final density matrix which determines whether the entropy shows non-monotonic behaviour in time. For non-Markovian dynamics, we show that the entropy approaches its steady state value as a power law with exponents governed by non-analyticities of the bath. We illustrate that this field-theoretic method can be used to study large bosonic open quantum systems.

 \end{abstract}
\title{Non-equilibrium Dynamics of Renyi Entropy for Bosonic Many-Particle Systems}
\author{Ahana Chakraborty}\email{ahana@pks.mpg.de}
 \affiliation{Department of Theoretical Physics, Tata Institute of Fundamental
 Research, Mumbai 400005, India.}
  \affiliation{Max Planck Institute for the Physics of Complex Systems, N\"othnitzer Str. 38, 01187, Dresden, Germany.}
\author{Rajdeep Sensarma}
 \affiliation{Department of Theoretical Physics, Tata Institute of Fundamental
 Research, Mumbai 400005, India.}

\date{\today}

\maketitle

A quantum state of a many body system encodes phase relations between spatially separated degrees of freedom. When this distributed information is erased by tracing out a set of degrees of freedom, the remaining subsystem is described by a reduced density matrix, which mixes quantum probability amplitudes with classical probabilities~\cite{Sakurai}. The Renyi entanglement entropy of the subsystem is the entropy of the resultant classical probability distribution. It indicates non-separability of the quantum state between the traced out and remaining degrees of freedom~\cite{Entanglement_Review}.
Renyi entropy of subsystems have been used to study quantum phase transitions~\cite{Renyi_QPT1,Renyi_QPT2,Renyi_QPT3} and many-body localization transition~\cite{EE_MBL1,EE_MBL2,EE_MBL3,EE_MBL4,EE_MBL5} in interacting disordered systems. It has been measured experimentally in ultracold atomic systems~\cite{Islam_expt}.

The simplest method for calculating entanglement entropy numerically diagonalizes the many body Hamiltonian~\cite{num_Ent1,num_Ent2}. While this works for fermions and spin systems of reasonable sizes~\cite{num_ent_large}, the large local Hilbert space of Bosons makes this method useless, unless one imposes hard-core constraints to whittle down the Hilbert space~\cite{ent_hardcore}. The problem becomes harder when one considers open quantum systems where number of particles is not conserved, and hence one cannot consider a truncated Hilbert space. Field theoretic methods, on the other hand, either use conformal invariance~\cite{Cardy_Ent_CFT,Cardy_Ent_CFT3,Renyi_CFT} or require correlation functions evaluated in replicated space-time sheets with branch-cuts along the sub-system~\cite{Cardy_Ent_CFT2}. In this Letter, we present a new method for calculating Renyi entropy of Bosonic many body systems, which can work without complicated manifolds and in absence of conformal invariance. Our method is applicable to systems both in and out of thermal equilibrium. It can describe evolution of Renyi entropy in non-Markovian open quantum systems starting from arbitrary initial conditions.

The Renyi entropy of a density matrix can be written as an integral of the square of its  Wigner quasiprobability distribution (WQD)~\cite{GlauberCahill}, which  is the closest approximation to a ``phase-space distribution function'' for quantum systems~\cite{Wigner1932,Wigner1984,GlauberCahill}.
The Wigner function has been measured in different single mode quantum systems~\cite{Faridani1993,Haroche2002,Mukamel1995,Wineland1996,Wigner_expt1,Wigner_expt2}.
In this letter, we show that the Wigner characteristic function (WCF) of a reduced density matrix is equal to the  partition function of the full system in presence of  a particular set of sources turned on only in the subsystem.
For systems out of thermal equilibrium, we use Schwinger Keldysh (SK) field theory to calculate this partition function, with recently proposed modifications to take care of arbitrary initial conditions~\cite{chakraborty2}.
The Wigner characteristics can then be used to calculate the Renyi entropy. We note that our formalism is quite different from earlier attempts to calculate WQDs~\cite{Anatoli_TWA,Anatoli_TWA1,Wigner_PI1,Wigner_PI2,Hofer_KQPD} within path integral approaches.

We use this formalism to study the evolution of Renyi entropy of Bosonic systems coupled to external baths, which are initialized to a Fock state with $0$ initial entropy. For a single mode system coupled to Markovian as well as non-Markovian baths~\cite{chakraborty1}, the entanglement entropy shows non-monotonic evolution in time in some cases. We find an explicit relation between the initial particle number and final density which determines whether the entropy evolution is non-monotonic. To illustrate the power of this new method, we then study the evolution of Renyi entropy in a linear chain of $ 12$ sites coupled to external non-Markovian baths. We recover the non-monotonic time dependence of the Renyi entropy of the subsystems.
We have also used this method to calculate Renyi entropies of large interacting thermal systems of Bosons in one and two dimensions, which is presented in a separate paper~\cite{chakrabortyrenyi}. Thus, our method can be used to calculate Renyi entropy of Bosonic many body systems in versatile circumstances not accessible by existing methods ~\cite{Ent_nonmarkov5,RenyiQme1,RenyiQme2,Zhang1,Zhang2,Zhang3,Zhang4,Sanz_2003}. 

\begin{figure}[t]
\includegraphics[width=0.49\textwidth]{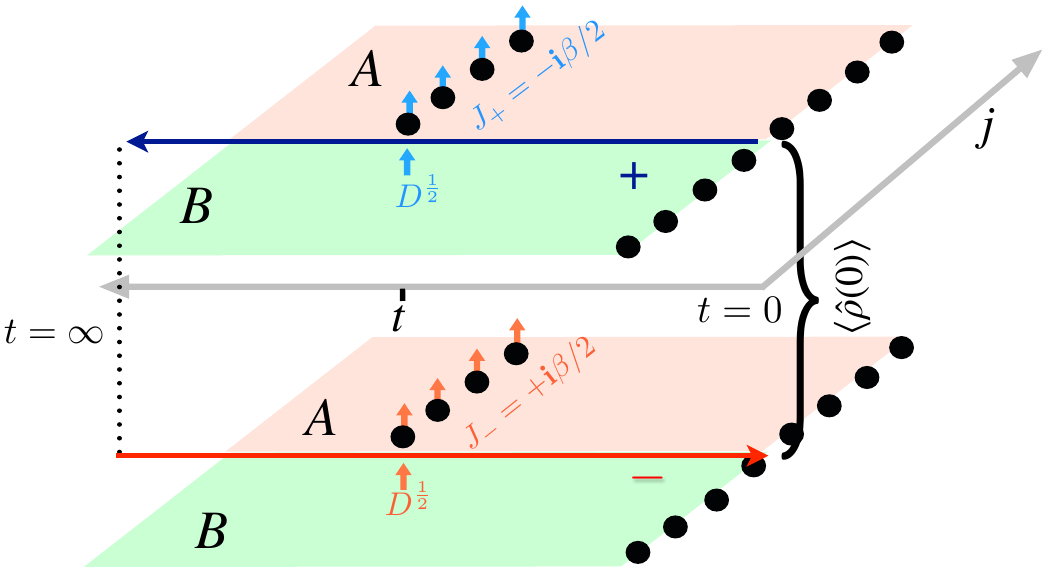}
\caption{Two contour ($\pm$) evolution of the density matrix, $\hat{\rho}(t)$ in SK field theory. For calculating $\chi_W(\beta_j,t)$, the displacement operator, $D(\{\beta_j\})$  is decomposed into $D \rightarrow D^{1/2}D^{1/2}$ with each $D^{1/2}$ inserted on the $+$ and $-$ contour at time $t$. For the WCF $\chi^A_W(\beta_j,t)$ of the reduced density matrix in subsystem $A$, this insertion is equivalent to turning on sources $J_+(j)=-\mathbf{i} \beta_j/2$ (shown by blue upward arrows) and $J_-(j)=+\mathbf{i} \beta_j/2$ (shown by orange upward arrows) only at the time of measurement $t$, and only on the lattice sites $j \in A$. The resultant partition function gives the WCF $\chi^A_W(\beta_j,t)$ from which $S^{(2)}(t)$ can be calculated using Eq.~\ref{S2_def}.} 
\label{fig:1}
\end{figure}

\begin{figure*}[t]
  \centering
  \includegraphics[width=0.7\textwidth]{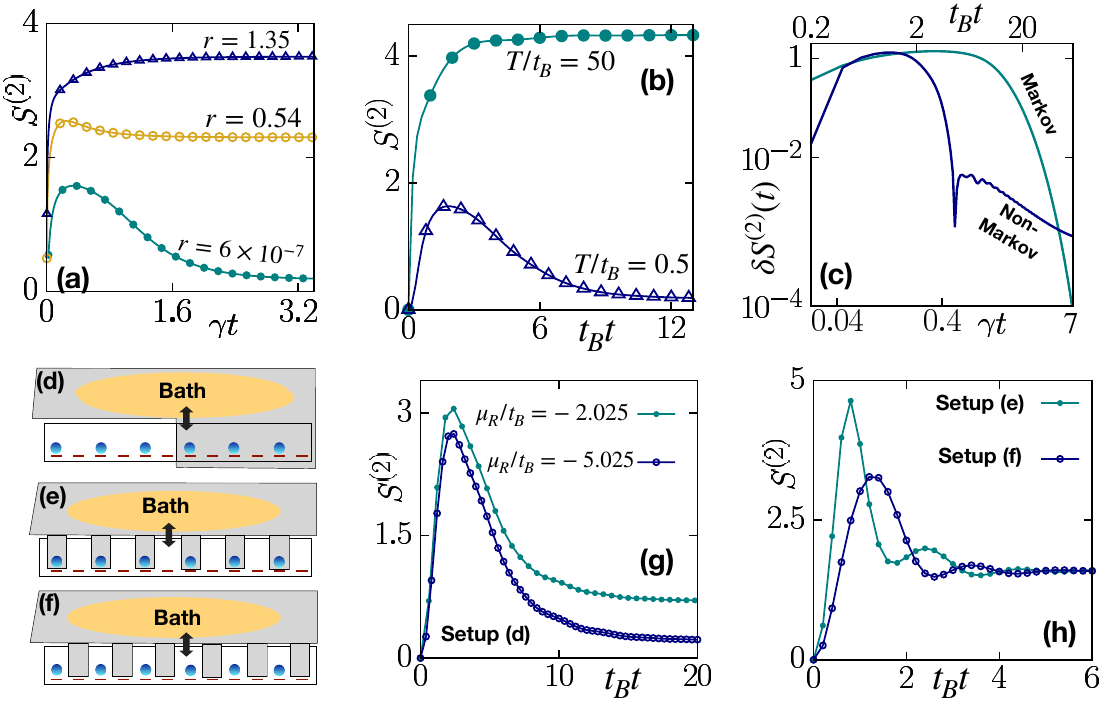} 
  \caption{ (a)-(b) Evolution of $S^{(2)}$ of a 1 mode system starting from $|n_i=6\rangle$ coupled to (a) a Markovian bath with increasing final number $n_f=0.1$, and $r=6 \times 10^{-7}$, $n_f=4.5$ and $r=0.54$,  $n_f=16.2$ and $r=1.35$. Here $r =n_f^{n_i}/(1 + n_f)^{(n_i + 1)} (1+2n_f) $; when $r<1$, $S^{(2)}$ shows non-monotonic behaviour in time. (b) a non-Markovian bath with $T/t_B=0.5$, showing a peak in $S^{(2)}$ and $T/t_B=50$ showing monotonic evolution of $S^{(2)}$. In general increasing $n_f$ with respect to $n_i$ leads to disappearance of peaks in $S^{(2)}(t)$. (c) $\delta S^{(2)}(t)= |S^{(2)}(t)-S^{(2)}(\infty)|$ as a function of $t$ in log-log plot. $S^{(2)}$ approaches to its steady state value exponentially for a Markovian bath and as a power law ($\sim t^{-3/2}$) for non-Markovian bath. (d)-(f) schematic representation of $12$-site linear chain initialized to $|101010...\rangle$ coupled to external bath. The grey shaded regions are integrated out to compute the Renyi entropy. (g) $S^{(2)}(t)$  for the setup of (d) where the right half of the system is integrated out. The two graphs correspond to a steady state with a finite current ($\mu_L/t_B=-2.025,\mu_R/t_B=-5.025$) and with zero current ($\mu_L/t_B=\mu_R/t_B=-2.025$). (h) $S^{(2)}(t)$ for the setups of (e) and (f) where the subsystem initialized to $1$ and $0$ particles in the initial state is integrated out respectively. $S^{(2)}(t)$ shows damped out of phase oscillations as the density imbalance between the sub-lattices is wiped out by the bath. Unless otherwise mentioned, the external bath has $T/t_B=0.5$ and $\mu/t_B=-2.025$. The non-Markovian bath has a bandwidth $t_B=2$ and a system bath coupling $\epsilon=1$. }
  \label{fig:onemode}
  \end{figure*}
{\bf Wigner Function, Renyi Entropy and Keldysh Partition Function:} For a many body bosonic system,  the WCF $\chi_W$ is given by~\cite{GlauberCahill}.
\begin{eqnarray}
\label{chiW_def}
\chi_W(\{\beta_j\},t)&=& Tr \left[\hat{\rho}(t) e^{\sum_j \beta_j a^\dagger_j 
  -\beta^\ast_j a_j}\right]
\end{eqnarray}
where $a^\dagger_j$ is the creation operator for the $j^{th}$ site, and $\hat{\rho}(t)$ is the density matrix of the system. The reduced density matrix of the subsystem $A$ (of size $\Omega_A$), $\hat{\rho}_{A}(t)=Tr_B \hat{\rho}(t)$ is obtained by tracing over the sites in the rest of the system ( $B$ ). The WCF $\chi_W^A$ of $\hat{\rho}_{A}(t)$ can be obtained from Eq.~\ref{chiW_def} by restricting the $\beta_j$s to the sites in $A$. The second Renyi entropy $S^{(2)} =- \log Tr\left[ \hat{\rho}_{A}^{2} (t) \right]$ is ~\cite{GlauberCahill}
\begin{equation}
\label{S2_def}
S^{(2)}(t) =- \log \left[\int\prod_{i \in A}\frac{d^2\beta_i}{\pi^{\Omega_A}}|\chi^{A}_W( \{ \beta_i\},t)|^2\right] .
\end{equation}
The time evolution of the density matrix, $\hat{\rho}(t)=U(t,0)\hat{\rho}(0) U^\dagger(t,0)$, can be described by a SK path integral with two copies of fields, $\phi_+(j,t)$ and $\phi_-(j,t)$, corresponding to the forward and backward evolution in time,  shown in Fig.~\ref{fig:1}. We decompose the displacement operator $D(\{\beta_j\})=e^{\sum_j \beta_j a^\dagger_j 
  -\beta^\ast_j a_j}$ in Eq.~\ref{chiW_def}, into $D\rightarrow D^{1/2} \times D^{1/2}$, with each of the $D^{1/2}$ placed on the $+$ and $-$ contour; i.e.
\begin{equation}
\chi_W= Tr \left[ U(\infty,t) D^{1/2}U(t,0)\hat{\rho}(0) U^\dagger(t,0) D^{1/2}U^\dagger(\infty,t) \right]
\end{equation}
 The insertion of $D^{1/2}=e^{\frac{1}{2}\sum_j \beta_j a^\dagger_j -\beta^\ast_j a_j}$ corresponds to turning on sources
$J_+(j,t')=-\mathbf{i}(\beta_j/2) \delta(t-t')$ and $J_-(j,t')=\mathbf{i}(\beta_j/2)
\delta(t-t')$ coupled linearly to the fields, and evaluating the SK partition function in presence of these sources [see Appendix A~\cite{note1} ].  Working with the classical and quantum fields, $\phi_{cl}=(\phi_++\phi_-)/\sqrt{2}$ and $\phi_{q}=(\phi_+-\phi_-)/\sqrt{2}$, the WCF is the Keldysh partition function $Z$ with the classical source turned off throughout the evolution, 
and a quantum source $J_{q}(j,t')=[J_+(j,t')-J_-(j,t')]/\sqrt{2}=-\mathbf{i}\beta_j/\sqrt{2}
\delta(t-t')$, turned on only at the time of measurement, i.e.
%
\begin{eqnarray}
\label{chiw_Z}
&&  \chi_W(\{\beta_j\},t) = Z\left[J_{cl}=0, J_q(j,t') =-\mathbf{i}\frac{1}{\sqrt{2}}\beta_j
\delta(t-t')\right]. ~~~~
\end{eqnarray}
This is the main result of this Letter, which is valid for generic non-equilibrium dynamics of interacting bosons.

In this formalism, integrating out modes without turning on a source traces over those degrees of freedom, while integrating out a mode after turning on a quantum source calculates the WCF. Hence $\chi_W^A$ can be obtained by restricting the quantum sources $\beta_j$ only to the modes in subsystem $A$ (See Fig.~\ref{fig:1} ). This simplification allows us to study Renyi entropy for arbitrary geometry of sub-systems.

{\bf Renyi Entropy in non-equilibrium dynamics:}
We consider a lattice system of Bosons initialized in a Fock state and coupled to an external bath. It evolves to a long time steady state under a non-unitary dynamics.

This dynamics can be treated within a recent extension of SK field theory \cite{chakraborty2}. For an initial $\hat{\rho}(0) = |\{n_i\}\rangle \langle \{n_i\}|$,
where $|\{n_i\}\rangle = |n_1,n_2 ...\rangle$, with $n_\nu$ the occupation number of the $\nu^{th}$ mode, the extended SK formalism adds to the Keldysh action a source $-\mathbf{i} (1+u_\nu)/(1-u_\nu)$ coupled to the bilinear $\phi^\ast_q(\nu,0)\phi_q(\nu,0)$, i.e. it is turned on only at the initial time $t=0$. The WCF $\chi_{W}(\beta_j,t,\vec{u})$ can then be calculated as the partition function in presence of both the linear quantum sources $\beta_j$, turned on at time of measurement $t$, and the initial bilinear sources $u_\nu$ turned on at $t=0$.
The physical WCF corresponding to the particular initial state is then obtained from
%
\begin{equation}
 \chi_W(\{\beta_j\},t) = \left. \prod_\nu \frac{1}{n_\nu!} \left(\frac{\partial}{\partial u_\nu} \right)^{n_\nu}\chi_W(\beta_j,t,\vec{u}) \right\vert_{u=0}.
\label{chi_W_rho}
\end{equation}
For a non-interacting system of bosons coupled to non-interacting bosonic baths,  $\chi_W(\beta_j,t,\vec{u})$, after integrating out the bath, is given by 
\begin{equation}
\chi_W(\beta_j,t,\vec{u}) =e^{-\frac{1}{2}\beta^\ast_i\beta_j\Lambda^0_{ij}(t)} \prod_\nu \frac{e^{-\beta^\ast_i\beta_j\Lambda^\nu_{ij}(t) \frac{u_\nu}{1-u_\nu}}}{1-u_\nu}
~~,
\end{equation}
where $\Lambda^0_{ij}(t) =\langle \phi_{cl}(i,t)\phi^\ast_{cl}(j,t)\rangle_0
$ is the equal time classical correlator in a system that starts from the vacuum state with $0$ particles, and $\Lambda^\nu_{ij}(t)=G^R_{i\nu}(t)G^{R\ast}_{j\nu}(t)$, where the retarded Green's function $G^R_{i\mu}(t)$ is the probability amplitude of finding a particle in mode $i$ at time $t$ if it was initially in mode $\mu$.  Our formalism can treat open and closed quantum systems, as well as Markovian and non-Markovian dynamics on equal footing. After taking the $\vec{u}$ derivatives \cite{grad}, the physical WCF is given by
\begin{equation}
\label{chi_W_Laguerre}
\chi_W(\beta_j,t) = e^{-\frac{1}{2} \beta^\ast_i\beta_j \Lambda^0_{ij}(t)} \prod_\nu L_{n_\nu}\left[ \beta^\ast_i\beta_j \Lambda^\nu_{ij}(t)\right]
\end{equation}
%
where $L_n(x)$ is the Laguerre polynomial of order $n$. 

The effect of the bath on the system dynamics is incorporated through the retarded self energy $\Sigma^R(i,j,\omega)$ (real part of $\Sigma^R$ controls the dressing of the system spectrum and its imaginary part controls the dissipation in the system) and the Keldysh self energy $\Sigma^K(i,j,\omega)$ (related to the stochastic noise from the bath and controls the steady state of the system). Inverting the Dyson equation, the correlators, and hence the evolution of the Renyi entanglement entropy can be computed [see Appendix B~\cite{note1}].

Note that this method can be easily generalized to interacting systems to start making approximations for entanglement entropy of interacting OQS (see Appendix A and C~\cite{note1}).

{\bf Single mode coupled to bath:} We first consider a single mode with $H=\omega_0 a^\dagger a$,  initialized in the number state $\hat{\rho}(0) =|n_i\rangle \langle n_i|$, and coupled to an external bath. The time-evolving Renyi entropy is given by,
\begin{equation}
e^{ -S^{(2)}(t)}= \! \frac{~{}^{2n_i}C_{n_i}~[\tilde{\Lambda}]^{2n_i}}{[\Lambda^0]^{2n_i+1}}{}_{2}F_1 \! \! \left[-n_i,-n_i,-2n_i,  \!\frac{-\Lambda^0[\Lambda-\tilde{\Lambda}]}{[\tilde{\Lambda}]^{2}} \! \right]
 \end{equation}
 where $\tilde{\Lambda}(t)=\Lambda^0(t)-\Lambda(t)$ and ${}_{2}F_1$ is the hypergeometric function. Here, as time increases $\Lambda(t)$ decays from $1$ to $0$ due to dissipation, while the stochastic contribution $\tilde{\Lambda}(t)$ increases from $0$ to a finite value determined by the bath parameters.

 The simplest bath is a Markovian Langevin type bath, characterized by a frequency independent $\Sigma^R(\omega)=\mathbf{i} \gamma$, where $\gamma$ is the dissipation scale and a frequency independent $\Sigma^K(\omega) =-\mathbf{i} D$ where $D$ is the noise scale. The steady state average number in the system is controlled by $1+2n_f =D/2\gamma$. The correlators in this case can be found analytically: $\Lambda^0(t) = e^{-2\gamma t} +\frac{ D}{2\gamma} [1-e^{-2\gamma t}]$ and $\Lambda(t) = e^{-2\gamma t}$.Fig~\ref{fig:onemode}(a) shows $ S^{(2)}(t)$ for this OQS starting from $n_i=6$ for three different values of $n_f=0.1,4.5$ and $16.2$. $ S^{(2)}(t)$ shows non-monotonic evolution with a peak at $t_{peak} \sim 1/2\gamma$ for $n_f=0.1$ and $n_f=4.5$, while the evolution is monotonic for $n_f=16.2$. A peak in $S^{(2)}$ implies that at $t\sim t_{peak}$, the initial delta function distribution of $\rho_{nn}$ spreads to a distribution which is wider than the final thermal distribution.

We next consider coupling the same system to a non-Markovian bath of bandwidth $t_B$
characterized by a spectral density $ \mathcal{J}(\omega) = \Theta(4t_B^2-\omega^2)\frac{2}{t_{B}} \sqrt{1-\frac{\omega^2}{4t^2_{B}}}$, temperature $T$ and chemical potential $\mu$ with coupling strength $\epsilon$~\cite{chakraborty1}. In this case, $\Sigma^R(\omega) = \mathbf{i} \epsilon^2 {\cal J}(\omega)$ and $\Sigma^K(\omega) = -\mathbf{i} \epsilon^2 {\cal J}(\omega) \coth [(\omega -\mu)/2T]$.
Fig~\ref{fig:onemode}(b) shows the evolution of  $S^{(2)}(t)$ for this OQS for $T/t_B=0.5$ where we see a non-monotonic time dependence with a peak at a timescale $\sim t_B/\epsilon^2$ and $T/t_B=50$ where the peak has disappeared. We note that contrary to previous predictions\cite{nonmonotonicPlenio,nonmonotonicMao}, the presence of the peak in $S^{(2)}(t)$ is not related to non-Markovian nature of the dynamics. The key difference between the Markovian and non-Markovian dynamics is manifested in how $S^{(2)}(t)$ approaches its steady state value. As shown in Fig~\ref{fig:onemode}(c), the Markovian approach is exponential while that for the non-Markovian case has a  power law tail ($t^{-3/2}$) at long times~\cite{chakraborty1}.

 To understand why the time evolution is non-monotonic in some cases, and monotonic in others, we use the ansatz $\hat{\rho}(t)= A(t) |n_i\rangle \langle n_i | + B(t) \hat{\rho}(\infty)$,
 where the first term denotes the forward time propagation (decay) of the initial state due to dissipation and the second term can be viewed as the backward time propagation of the steady-state density matrix of the system $\hat{\rho}(\infty)$. For the Langevin bath, we can obtain exact expressions, $A(t)=|G^R(t,0)|^2=e^{-2\gamma t}$ and $B(t)=1-e^{-2\gamma t}$. While similar expressions for the non-Markovian bath can only be written in terms of multiple integrals, it is easy to see that in both cases $A(t) \sim 1- at$ and $B(t) \sim b t$ for short times, where $a$ and $b$ are rates that can be computed from a Fermi golden rule calculation.  Also $A(t) \rightarrow 0$ and  $B(t) \rightarrow 1$ as $t\rightarrow \infty$ to obtain the correct steady state density matrix.  We can then obtain the time dependent Renyi entropy
 \begin{equation}
  e^{-S^{(2)}(t)}= A^2(t) + B^2(t) e^{-S^{(2)}(\infty)}+ 2 A(t) B(t)  \rho_{n_i,n_i}(\infty),~~~
\label{entgenform}
\end{equation}
where $ \rho_{n_i,n_i}(\infty)$ is the diagonal matrix element of $\hat{\rho}(\infty)$ in the initial state $|n_i\rangle$.

The initial rise of the entanglement is controlled by the rates $a$ and $b$ with $e^{-S^{(2)}(t)} \sim 1- 2at (1-\rho_{n_i,n_i}(\infty) b/a)$. Interestingly this rate is neither controlled by the initial state, nor the final density matrix, but a cross term between them. For an heuristic argument, we can compare the entropy  at a time $t_0\sim (2a)^{-1}$, $e^{-S^{(2)}(t_0)} \sim \rho_{n_i,n_i}(\infty) b/a$, with its steady state value $e^{-S^{(2)}(\infty)}$ and say that a peak occurs if   $\rho_{n_i,n_i}(\infty) b/a<e^{-S^{(2)}(\infty)}$. For the Langevin bath $b=a=2\gamma$, so this criterion reduces to  $\rho_{n_i,n_i}(\infty) <e^{-S^{(2)}(\infty)}=(1+2 n_f)^{-1}$, which gives $r<1$, where $r =n_f^{n_i}/(1 + n_f)^{(n_i + 1)} (1+2n_f)$. We also find the peak condition for the Langevin bath from an exact calculation involving maxima of Eq.~\ref{entgenform} (see Appendix D~\cite{note1}) and get the same result. Increasing average density in a thermal system leads to a broader distribution of $\rho_{nn}(\infty)$, and hence a decrease in $exp[-S^{(2)}(\infty)]$. Thus, a general conclusion from this is that increasing $n_f$ keeping $n_i$ fixed leads to disappearance of the peak in $S^{(2)}$.


  {\bf Entanglement Dynamics of many body system:} For a single or two mode system the dynamics of Renyi entropy can in principle be studied by more direct means like quantum master equations ~\cite{Ent_nonmarkov5,RenyiQme1,RenyiQme2,Zhang1,Zhang2,Zhang3,Zhang4,Sanz_2003}. However, these methods are not easy to generalize to many body Bosonic systems due to the infinite size of the local Hilbert space. Our method, however, is easily generalisable to dynamics of many body open quantum systems initialized to non-thermal states. The truncation of the Hilbert space in the existing methods is specially problematic for open quantum systems where the particle number is not conserved. While methods like TEDOPA chain mapping~\cite{Priortedopa} try to get around this by working with a finite sized bath, our method can easily incorporate infinite baths. 
  
  To show the power of this method, we consider a linear chain of Bosons described by the Hamiltonian
  \begin{equation}
   H=-g \sum_{j} a^{\dagger}_{j+1} a_j+h.c., 
\end{equation}  
with the nearest neighbour hopping $g=0.5 t_B$. This system is made of the even ($E$) and odd ($O$) sublattices and is initialized to a Fock state with $0$ and $1$ number of particles on the $E$ and $O$ sublattices respectively (see Fig.~\ref{fig:onemode}(d)-(f)). Each site of the system is coupled to the non-Markovian bath described earlier. The temperature of the baths is set to $T/t_B=0.5$.
We treat two different cases: (a) when the baths have the same $T$ and $\mu$, leading to thermalization at long times, and (b) when $\mu$ of the baths coupled to neighbouring sites have a fixed difference of chemical potential, leading to steady states with finite currents.
In this case, 
\begin{equation}
 e^{-S^{(2)}(t)}=\! \int \! \prod_{i\in A}\frac{d^2 \beta_i}{\pi^{\Omega_A}} \! e^{- \sum \limits_{i,j \in A}\beta^\ast_i\beta_j \Lambda^0_{ij}(t)} \! \! \prod_{\nu\in O}\! L^2_{1}\left[ \beta^\ast_i\beta_j \Lambda^\nu_{ij}(t)\right],
  \end{equation}
  where $A$ denotes the reduced sub-system and $O$ denotes the initially filled sub-lattice~\cite{note}.

 In a many body system, we can trace out a part of the system, together with the bath, to construct the subsystem for which we consider Renyi entropy. Fig.~\ref{fig:onemode}(g) shows evolution of $S^{(2)}(t)$ of the subsystem which occupies the left half of the linear chain (see setup in Fig.~\ref{fig:onemode}(d)). The chemical potential for the leftmost bath is set to be $\mu_L/t_B=-2.025$, so the graph for $\mu_R/t_B=-2.025$ corresponds to identical baths, while the graph for $\mu_R/t_B=-5.025$ corresponds to a linear gradient in $\mu$ resulting in a finite current in the steady state. We find that $S^{(2)}(t)$ is non-monotonic in both cases, and the steady state value is lower for the system with a finite current. In Fig.~\ref{fig:onemode}(h), we consider the same system, with the same initial condition of alternating filled and empty sites, but our subsystem is now composed of the $A$ ($B$) sublattice obtained by integrating out the $B$ ($A$) sublattice, as shown in Fig.~\ref{fig:onemode}(e) and (f). The evolution of Renyi entropy shows damped out-of-phase oscillation in $S^{(2)}(t)$ between the two setups, which track the oscillations in the imbalance of density on the two sublattices. The steady state value is same in both cases, showing that the system erases information about initial state.

We have developed a new method for calculating Renyi entropy of Bosonic many body systems in and out of thermal equilibrium. 
This method has also been used to study Renyi entropy of thermal system of interacting Bosons in one and two dimensions of large sizes within a large $N$ approximation~\cite{chakrabortyrenyi}. These ideas can also be extended to the case of Fermionic systems with appropriate modifications\cite{saranyo,sumilan}.

\begin{acknowledgments}
The authors thank Anatoli Polkovnikov and Takashi Oka for useful discussions. The authors also acknowledge computational facilities at Department of Theoretical Physics, TIFR Mumbai. 
\end{acknowledgments}

\bibliographystyle{apsrev4-1}
\bibliography{main.bib}

\begin{thebibliography}{54}
\bibitem [{not({\natexlab{b}})}]{note}%
  \BibitemOpen
  \href@noop {} {} ({\natexlab{b}}),\ \bibinfo {note} {here we have worked with
  a linear chain of $12$ sites. In existing numerical methods this would have
  corresponded to Hilbert space of size $\sim 10^8$, even if the local Hilbert
  space is truncated at maximum number state of $|n_{max}\rangle=|4 \rangle$,
  and hence is beyond the capacity of current numerical methods.}\BibitemShut
  {Stop}%
\end{thebibliography}

\end{document}


\newcommand {\beq} {\begin{equation}}
\newcommand {\eeq} {\end{equation}}
\newcommand {\bqa} {\begin{eqnarray}}
\newcommand {\eqa} {\end{eqnarray}}
\newcommand {\ba} {\ensuremath{b^\dagger}}
\newcommand {\Ma} {\ensuremath{M^\dagger}}
\newcommand {\psia} {\ensuremath{\psi^\dagger}}
\newcommand {\psita} {\ensuremath{\tilde{\psi}^\dagger}}
\newcommand{\lp} {\ensuremath{{\lambda '}}}
\newcommand{\A} {\ensuremath{{\bf A}}}
\newcommand{\Q} {\ensuremath{{\bf Q}}}
\newcommand{\kk} {\ensuremath{{\bf k}}}
\newcommand{\qq} {\ensuremath{{\bf q}}}
\newcommand{\kp} {\ensuremath{{\bf k'}}}
\newcommand{\rr} {\ensuremath{{\bf r}}}
\newcommand{\rp} {\ensuremath{{\bf r'}}}
\newcommand {\ep} {\ensuremath{\epsilon}}
\newcommand{\nbr} {\ensuremath{\langle ij \rangle}}
\newcommand {\no} {\nonumber}
\newcommand{\up} {\ensuremath{\uparrow}}
\newcommand{\dn} {\ensuremath{\downarrow}}
\newcommand{\rcol} {\textcolor{red}}

\title{Supplementary Material for "Non-equilibrium Dynamics of Renyi Entropy for Bosonic Many-Particle Systems"}
\author{Ahana Chakraborty}\email{ahana@pks.mpg.de}
 \affiliation{Department of Theoretical Physics, Tata Institute of Fundamental
 Research, Mumbai 400005, India.}
  \affiliation{Max Planck Institute for the Physics of Complex Systems, N\"othnitzer Str. 38, 01187, Dresden, Germany.}
\author{Rajdeep Sensarma}
 \affiliation{Department of Theoretical Physics, Tata Institute of Fundamental
 Research, Mumbai 400005, India.}

\pacs{}
\date{\today}

\maketitle

\appendix

\section{WCF from Keldysh partition function with particular choice of classical and quantum sources} 
%
%

In this appendix, we will detail the idea of relating the WCF of a Bosonic system with Keldysh partition function in presence of a particular combination of the linear sources in the two-contour Keldysh action. In SK field theoretic approach, the partition function $Z$, which encodes the time evolution of the many body density matrix, $\hat{\rho}(t)=U(t,0)\hat{\rho}(0) U^\dagger(t,0)$ along the closed contour as shown in Fig.1 of the main text, is expressed as the path integral over $\phi_+(j,t)$ and $\phi_-(j,t)$ fields which are linearly coupled to the sources, $J_+(j,t)$ and $J_-(j,t)$.
\begin{widetext}
\begin{equation}
\label{AP:Z}
Z\left[ J_+,J_- \right] = \int D[\phi_+]D[\phi_-] e^{\mathbf{i}\left[ \mathcal{A}_+ - \mathcal{A}_- \right] +\mathbf{i}\sum \limits_j \int ~dt~ \left[J_+(j,t) \phi_+^\ast(j,t) +h.c. -J_-(j,t) \phi_-^\ast(j,t) -h.c. \right ]},
\end{equation}
\end{widetext}
where $D[\phi] = \prod_n d\phi^\ast(t_n)d\phi(t_n)/\pi$ and $\mathcal{A}_{\pm}$ encodes the evolution of $\hat{\rho}(t)$ along the $\pm$ branches. For a non-interacting closed system, governed by the Hamiltonian $H_s=\sum_{i,j}h_{ij} a^\dagger_i a_j + h.c.$, $\mathcal{A}_{\pm} = \sum_{j,j'}\int~dt \int~dt'[ \phi_{\pm}^\ast(j,t)(i\partial_t-h_{ij})\phi_{\pm}(j't')]$, whereas for an OQS coupled linearly to an external bath, $\mathcal{A}$ will be modified by the corresponding bath induced self-energies. We note that the information about the initial density matrix of the system is incorporated as a boundary term in the action $\mathcal{A}$ ~\citep{kamenev} which we will discuss in the next section in great details.

 To calculate the expectation value,$\langle \mathcal O \rangle$ of any observable, $\mathcal{O}$, the standard technique in Keldysh field theory is to additively decompose the operator into $\mathcal O = (\mathcal O_+ + \mathcal O_-)/2$ and write down the path integral for $\langle\mathcal O_+\rangle (\langle\mathcal O_-\rangle)$ by inserting the operator only on the forward (backward) branch of the Keldysh contour. Taking derivatives of $Z\left[ J_+,J_- \right] $ w.r.t the sources $J_+$ and $J_-$ accordingly gives the corresponding contributions, $\langle \mathcal O_+ \rangle$ and  $\langle \mathcal O_- \rangle$, from $+$ and $-$ branches respectively. Finally the average of these two contributions yields the quantum mechanical average of the physical observable in the density matrix $\hat{\rho}(t)$.

In this letter, in order to calculate WCF, $\chi_W(\{\beta_j\},t)= Tr \left[\hat{\rho}(t) D\right]$, we propose a different technique to calculate the expectation value of the displacement operator, $D(\{\beta_j\})=e^{\sum_j \beta_j a^\dagger_j-\beta^\ast_j a_j}$, where the creation and annihilation operator already appear in the exponent. Instead of the additive decomposition of the operator stated above, we perform a multiplicative decomposition of the displacement operator, $D\rightarrow D^{1/2}_+ \times D^{1/2}_-$, and insert $D^{1/2}$ symmetrically on both the branches within a single two contour evolution represented by the path integral,
\begin{widetext}
\beq
\label{AP:chiW}
\chi_W= Tr \left[  U^\dagger(t,0) ~e^{\frac{1}{2}\sum_j \beta_j a^\dagger_j-\beta^\ast_j a_j}~U^\dagger(\infty,t)   U(\infty,t)~e^{\frac{1}{2}\sum_j \beta_j a^\dagger_j-\beta^\ast_j a_j}~U(t,0) \hat{\rho}(0) \right].
\eeq
\end{widetext}
Here it is important to note that $D^{1/2}$ is not a normal-ordered operator, but in the path integral equation \ref{AP:chiW}, the commutation relations between $a_i,a^{\dagger}_j$, in the expansion $e^{\frac{1}{2}\sum_j \beta_j a^\dagger_j-\beta^\ast_j a_j}$ are exactly captured by the commutation relations of the operator valued fields $\phi_{\pm},\phi_{\pm}^{\ast}$ at equal times. 
Comparing equation \ref{AP:chiW} with \ref{AP:Z}, we observe that at any instant of time $t$, $\chi_W(\{\beta_j\},t)$ is equal to the Keldysh partition function $Z\left[ J_+,J_- \right] $ if the linear source fields $J_+,J_-$ are identified with the delta-function sources turned on only at the instant $t$, i.e. 
\beq
J_\pm(j,t')=\mp \mathbf{i}(\beta_j/2) \delta(t-t').
\eeq
The identification of $\chi_W = Z[J_\pm(j,t')=\mp \mathbf{i}(\beta_j/2) \delta(t-t')]$ becomes further simplified if we work in the rotated $cl/q$ basis, where $J_{cl}(j,t')=[J_+(j,t')+J_-(j,t')]/\sqrt{2}$ and $J_{q}(j,t')=[J_+(j,t')-J_-(j,t')]/\sqrt{2}$,
\beq
\chi_W(\{\beta_j\},t) = Z\left[J_{cl}=0, J_q(j,t') =-\mathbf{i}\frac{1}{\sqrt{2}}\beta_j
\delta(t-t')\right] .
\label{rel_chiWZ}
\eeq
At this point, we would like to point out that the choice of placing $D^{1/2}$ symmetrically on both the branches and hence setting $J_\pm(j,t')=\mp i(\beta_j/2) \delta(t-t')$ is not a unique choice. But it is the most convenient choice of the sources as it leads to $J_{cl}(j,t) = 0$ for all time and  $J_q(j,t') =\mathbf{i}\beta_j
\delta(t-t')/\sqrt{2}$ couples only to $\phi_{cl}(j,t)$. Hence $\chi_W(\{\beta_j\},t)$ (and also $W(\{\alpha_j\},t), S^{(2)}(t)$) for a generic many body (interacting) system can be completely formulated in terms of only the classical correlation functions at equal time. For a  non-interacting (open quantum systems) starting from thermal states, the Gaussian integration in equation \ref{AP:Z} can be performed exactly to obtain,
\begin{eqnarray}
\nonumber\chi_{W}(\beta_j,t) = e^{-\frac{1}{2} \beta^\ast_i \beta_j \Lambda_{ij}(t)} ~~~~,~~~ W(\alpha_j,t) = \frac{e^{-\frac{1}{2} \alpha^\ast_i \alpha_j \Lambda^{-1}_{ij}(t)} }{Det [\hat{\Lambda}/2]}~~~~,~~~
 S^{(2)} =Tr~ log \Lambda_{ij}(t)  ,
 \label{chi_W_NI} 
\end{eqnarray} 
where the one particle Keldysh correlator $\Lambda_{ij}(t) =\langle \phi_{cl}(i,t) \phi_{cl}^\ast(j,t)\rangle$ contains the effects of the bath (see Refs.~\onlinecite{Chakraborty1, Nori}). The WCF and WQD are both gaussian in this case, while the Renyi entropy matches with earlier results~\cite{Casini_Ent_QFT}. 
 The modes $i,j$ belong to the remaining Hilbert space $A$, which could be the full system (bath integrated out) or a part of the system ( bath +part of system integrated out). 
For an interacting system, we use the fact that the logarithm of a partition function is the generator of connected correlators to write 
\begin{equation}
\chi_{W}(\beta_j,t) = e^{\sum \limits_n (-1)^n\frac{1}{2^n n!^2}\Lambda^{i_1..i_n}_{j_1..j_n}(t) \beta^\ast_{i_1}.. \beta^\ast_{i_n} \beta_{j_1}.. \beta_{j_1}}
\label{chiw_int}
\end{equation}
where $\Lambda^{i_1..i_n}_{j_1..j_n}(t) = \langle \phi_{cl}(i_1,t) .. \phi_{cl}(i_n,t) \phi^\ast_{cl}(j_1,t) .. \phi^\ast_{cl}(j_n,t)\rangle_C$ is the connected $n$ particle equal time correlator of the classical fields for $n\neq 1$ and for $n=1$, $\Lambda^{i}_j=\Lambda_{ij}$. This exact relation between $\chi$ (and hence $S^{(2)}$) and the standard correlation functions is the starting point for making different approximations for entanglement entropy of interacting systems.

\section{Construction of equal time classical correlator $\Lambda(t)$ for OQS with non-thermal $\hat{\rho}(0)$:}
In the main text of this letter, we have shown that the WCF of a Bosonic system is the SK partition function of the system in presence of a particular set of sources.
We now present the details of calculating this partition function for a non-interacting system starting from an arbitrary Fock state. We note that this requires an extension of the standard SK field theory recently developed by some of us \cite{chakraborty2}. We will repeat some of the algebra for the sake of completeness here.

We consider a many body system starting from an initial density matrix,
$
\hat{\rho}(0) = |\{n\}\rangle \langle \{n\}|
$ where
$|\{n\}\rangle =\prod_\nu |n_\nu \rangle$ is a configuration in the Fock space. The partition function in presence of the sources including the boundary term can be written as, 
%
\begin{widetext}
\beq
\label{AP:Z_rho}
Z_{\rho}\left[ J_+,J_- \right] = \int D[\phi_+]D[\phi_-] e^{\mathbf{i}\left[ \mathcal{A}_+ - \mathcal{A}_- \right] +\mathbf{i}\sum_j \int ~dt~ \left[J_+(j,t) \phi_+^\ast(j,t) +h.c. -J_-(j,t) \phi_-^\ast(j,t) -h.c. \right ]} \langle \phi_+(0)| \hat{\rho}(0)|\phi_-(0)\rangle.
\eeq
\end{widetext}
Here, the matrix element of $\hat{\rho}(0)$ can be written as a term in the action where a source $u_\nu$, turned on only at $t=0$, is coupled to the bi-linears of the initial fields $ \phi_+(\nu,0)\phi_-(\nu,0)$. This is done by the virtue of the identity,
\beq
\langle \phi_+(0)| \hat{\rho}(0)|\phi_-(0)\rangle=
\left.\prod_\nu\frac{1}{n_\nu!}\left [
  \frac{\partial}{\partial {u_\nu}}\right]^{n_\nu} e^{\sum_\kappa u_\kappa
  \phi^\ast_{+}(\kappa, 0)\phi_{-}(\kappa,0)}\right\vert_{\vec{u}=0} .\nonumber
\eeq 
At this point, we would like to note the difference between the quadratic source $u_\nu$, turned on only at the initial time $t=0$, and the linear sources, $J_\pm(j,t')=\mp i(\beta_j/2) \delta(t-t')$, turned on only at the time measurement $t$ of WQD. Here, the role of the quadratic source, $\vec{u}$ is to incorporate the initial non-thermal density matrix, whereas the linear sources $J_{\pm}(j,t)$ are turned on to measure $\chi_W(\{ \beta_j\},t)$ at time $t$.

The physical partition function, $Z_{\rho}[J_+(j,t),J_-(j,t)]$ corresponding to the initial density matrix, $\hat{\rho}(0)$ is computed in a two step process: first we will calculate the partition function $Z[J_+(j,t),J_-(j,t),\vec{u}]$ in presence of the initial source $\vec{u}$ and then by taking its appropriate number of derivative w.r.t $\{u_{\nu}\}$, dictated by $\hat{\rho}(0)$, we will get $ Z_{\rho}$ given by equation \ref{AP:Z_rho}. The calculation of WCF also follows the two step process stated above for the partition function: first we will calculate 
\beq
\label{AP:chi_W_u}
\chi_{W}(\beta_j,t,\vec{u})= Z[J_\pm(j,t')=\mp i(\beta_j/2) \delta(t-t'),\vec{u}]
\eeq
 followed by 
 \beq
 \label{AP:chi_W_deriv}
  \chi_W(\{\beta_j\},t) = \left. \prod_\nu \frac{1}{n_\nu!} \left(\frac{\partial}{\partial u_\nu} \right)^{n_\nu}\chi_W(\beta_j,t,\vec{u}] \right\vert_{u=0}.
\eeq
For a non-interacting closed quantum system, governed by the Hamiltonian $H_s=\sum_{i,j}h_{ij} a^\dagger_i a_j + h.c.$, it has been shown in reference \onlinecite{chakraborty2} that the partition function in presence of the initial source $\vec{u}$, can be written in the $cl/q$ basis , 
\begin{widetext}
\beq
Z[J_{cl}(j,t),J_{q}(j,t),\vec{u}] =\int D[\phi_{cl}]D[\phi_q]~e^{\left[\mathbf{i}\int_0^\infty dt \int_0^\infty dt'\sum \limits_{j,j'} \phi^{\dagger}(j,t) \hat{G}^{-1}(j,t;j',t')
\phi(j',t') +\mathbf{i} \sum \limits_j \int_0^\infty dt J^{\dagger}(j,t) \phi(j,t) + h.c\right]} ,
\eeq
\end{widetext}
where $\phi^\dagger(j,t)=[\phi^\ast_{cl}(j,t),\phi^\ast_{q}(j,t)]$ and $J^\dagger(j,t)=[J^\ast_{cl}(j,t),J^\ast_q(j,t)]$.
The explicit $\vec{u}$ dependence is shown in the inverse Green's function as,
\bqa
&&\hat{G}^{-1}(j,t;j',t')=\left(\begin{array}{cc}
0& G_A^{-1}(j,t;j',t') \nonumber \\
G_R^{-1}(j,t;j',t')& G_K^{-1}(j,t;j',t',\vec{u})
\end{array}\right),\nonumber \\
&&\nonumber G_R^{-1}(j,j',t-t')=\delta(t-t')[ i\partial_t \delta_{j,j'}-h_{jj'}], \nonumber \\
&& G_K^{-1}(j,t;\nu,t',\vec{u})=-\mathbf{i}\delta_{j,\nu}\frac{1+u_\nu}{1-u_\nu} \delta(t)\delta(t').
\label{AP:inv_GF}
\eqa
Inverting the kernel in the action, we can write down the Keldysh Green's function in presence of the initial source $\vec{u}$ as \cite{chakraborty2},
\beq
\label{AP:Greens_GK}
G^K_{ij}(t,t',\vec{u})
=-\mathbf{i}\sum_{\nu}\frac{1+u_\nu}{1-u_\nu}
  G^{R}_{i\nu}(t)G^{R*}_{j\nu}(t'),
\eeq
whereas the retarded Green's function $G^{R}_{ij}(t-t')$ is independent of the initial source $\vec{u}$ and hence a function of $(t-t')$.

We will next extend the above formalism for an open quantum system. We consider a generic OQS model where a non-interacting system of Bosons ($H_s$) is linearly coupled to a non-interacting Bosonic bath $H_b$ via the particle exchange Hamiltonian $H_{sb}$ given by,
\beq
\label{AP:Ham_B}
H_b= \sum_{i,\alpha}\Omega_{\alpha} B^{(i)\dagger}_{\alpha} B^{ (i)}_{\alpha} ~,~
H_{sb}  =\epsilon \sum \limits_{i,\alpha} \kappa_{\alpha} B^{(i)\dagger}_{\alpha} a_{i} + h.c ,
\eeq
where $B^{(i)\dagger}_{\alpha}$ creates a Boson in the $\alpha^{th}$ mode of the $i^{th}$ bath and $\kappa_{\alpha}$ denotes the form system-bath coupling between the $i^{th}$ site of the system and the $\alpha^{th}$ mode of the bath with strength $\epsilon$. The system bath coupling is turned on at $t=0$. All the baths are assumed to be in thermal equilibrium with temperature $T_i$ and chemical potential $\mu_i$ throughout the dynamics. After tracing out bath degrees of freedom, we will study the reduced dynamics of the system where the effect of the bath comes in the form of the retarded and Keldysh self-energies, $\Sigma_R^B(i,j,t-t')$ and $\Sigma_K^B(i,j,t-t')$ respectively in the effective action of the system \cite{Chakraborty1}.
\bqa
\label{AP:eq_GF_bath}
G_{A}^{-1}(j,j',t-t')&=& G_{Av}^{-1}(j,j',t-t')-\Sigma_A^B (j,j',t-t') ,\nonumber \\
G_{R}^{-1}(j,j',t-t')&=& G_{Rv}(j,j',t-t')^{-1}-\Sigma_R^B(j,j',t-t') ,\nonumber \\
G_{K}^{-1}(j,t,j',t')&=& -\mathbf{i}\frac{1+u_\alpha}{1-u_\alpha}\delta_{j,j'}\delta_t\delta_{t'}-\Sigma_K^B(j,j',t-t'),\nonumber \\ 
\eqa
where $G_{Av}^{-1},G_{Rv}^{-1}$ are given by equation \ref{AP:inv_GF}. We use the subscript $v$ to denote a system starting from vacuum ($0$ particles) as the retarded and advanced component of the equation \ref{AP:eq_GF_bath} are independent of $\vec{u}$.  We note that the bath induced self-energies $\Sigma_R^B(j,j',t-t'),\Sigma_K^B(j,j',t-t')$ are controlled by the bath spectral function, $\mathcal{J}(\omega) = 2 \pi \sum_{\alpha} |\kappa_{\alpha}|^{2} \delta(\omega-\Omega_{\alpha})$ \citep{Nori,Chakraborty1} and the distribution function of the bath
\bqa
\label{AP:sig_t}
\Sigma_R^B(i,j,t-t') &=&-\mathbf{i} \delta_{i,j}~\epsilon^2\Theta(t-t') \int \frac{d\omega}{2\pi} \mathcal{J}(\omega)e^{-\mathbf{i}\omega(t-t')},\\
\nonumber\Sigma_K^B(i,j,t-t') &=&-\mathbf{i} \delta_{i,j}~\epsilon^2 \int \frac{d\omega}{2\pi} \mathcal{J}(\omega)\coth \left[\frac{\omega-\mu_i}{2T_i}\right] e^{-\mathbf{i}\omega(t-t')},
\eqa
and hence they are independent of the initial condition of the system encoded by the source, $\vec{u}$. 
$G^{R}_{ij}(t-t')$ is obtained by performing a modified Laplace transform in the \textit{complex} frequency space \cite{Nori}. Here we will use the analytical solution of the retarded Green's function, $G^{R}_{ij}(\omega)$ of a many body system provided in our earlier work in reference \onlinecite{Chakraborty1} where the system-bath coupling strength is assumed to be same for all sites, i.e. $\kappa_{\alpha}$ is independent of $i$ in equation \ref{AP:Ham_B}. On the other hand, the information about the initial condition of the system is carried to the subsequent dynamics by the Keldysh Green's function through an explicit dependence on $\vec{u}$ written as, 
\begin{widetext}
\beq
\label{AP:Greens_GK_B}
G^K_{ij}(t,t',\vec{u})
=-\mathbf{i} \sum_{\nu}\frac{1+u_\nu}{1-u_\nu}
  G^{R}_{i\nu}(t)G^{R*}_{j\nu}(t')+ \int_0^t
  dt_1\int_0^{t'} dt_2 \sum \limits_{m,n} G^{R}_{im}(t-t_1)\Sigma_K^B(m,n,t_1-t_2)G^{A}_{n j}(t_2-t').
\eeq
\end{widetext}
For further simplification of the calculation, it is useful to isolate the $\vec{u}$ dependent part of $G^K_{ij}(t,t',\vec{u})$ from the vacuum Keldysh Green's function as,
\beq
G^K_{ij}(t,t',\vec{u}) = G^K_{ij}(t,t',0) -\mathbf{i} \sum_{\nu}\frac{2 u_\nu}{1-u_\nu}
  G^{R}_{i\nu}(t)G^{R*}_{j\nu}(t').
\eeq
The one-particle classical correlation function in vacuum, which is defined as $\Lambda_{ij}^0(t)$ in the main text, is related to the equal time Keldysh Green's function by,
\begin{widetext}
\beq
\Lambda_{ij}^0(t) = \mathbf{i}G^K_{ij}(t,t,\vec{u}=0) = -\mathbf{i} \sum_{\nu}
  G^{R}_{i\nu}(t)G^{R*}_{j\nu}(t)+ \int_0^t
  dt_1\int_0^{t} dt_2 \sum_{m,n} G^{R}_{im}(t-t_1)\Sigma_K^B(m,n,t_1-t_2)G^{A}_{n j}(t_2-t) .
  \eeq.

Now in order to find out WCF for a non-interacting OQS in presence of $\vec{u}$, the functional integral integral over classical and quantum fields in the partition function can be done exactly to obtain equation \ref{AP:chi_W_u} in the following form,
\bqa
\chi_W(\beta_j,t,\vec{u}) =  \prod_\kappa \left(\frac{1}{1-u_\kappa}\right)exp\left[-\frac{1}{2}\beta^\ast_i\beta_j ~\mathbf{i} G^K_{ij}(t,t,\vec{u})\right] 
= \prod_\kappa \left(\frac{1}{1-u_\kappa}\right)exp\left[-\frac{1}{2}\beta^\ast_i\beta_j \left(\Lambda^0_{ij}(t) +\sum_\nu \Lambda^\nu_{ij}(t) \frac{2u_\nu}{1-u_\nu}) \right)\right] , \nonumber
\eqa
where $\Lambda^\nu_{ij}(t)=G^R_{i\nu}(t)G^{R\ast}_{j\nu}(t)$. This can be written as,
\beq
\chi_W(\beta_j,t,\vec{u}) = e^{[-\frac{1}{2}\beta^\ast_i\beta_j (\Lambda^0_{ij}(t) ]}\prod_\nu \left(\frac{1}{1-u_\nu}\right)exp^{\left[-\beta^\ast_i\beta_j \left( \Lambda^\nu_{ij}(t) \frac{u_\nu}{1-u_\nu}) \right)\right] }.
\eeq
Since $exp[-x u/(1-u)]/(1-u)$ is the generating function of the Laguerre polynomial $L_n(x)$, 
taking derivative of $\chi_W(\beta_j,t,\vec{u})$ w.r.t $u_\nu$, yields the physical WCF in terms of Laguerre polynomial given by equation 7 of the main text. At this point, it is worthwhile to mention that the case of a thermal initial density matrix, $\hat{\rho}(0)= exp\left [ -(\omega_\nu -\mu)a^{\dagger}_\nu a_\nu /2T \right]$ is obtained as special case of the formalism stated above. Instead of taking derivatives we just need to set the initial sources $u_\nu = exp\left[ -(\omega_\nu -\mu)/2T \right]$. The Green's functions in equation \ref{AP:Greens_GK} then recover their thermal form with $\Lambda_{ij} $ given by,

\bqa
\Lambda_{ij} &=& \mathbf{i}G^K_{ij}(t,t,u_\nu = e^{ -(\omega_\nu -\mu)/2T}) \nonumber \\
&=& -\mathbf{i} \sum_{\nu}\coth\left( \frac{\omega_\nu - \mu}{2T} \right)
  G^{R}_{i\nu}(t)G^{R*}_{j\nu}(t)+
  \int_0^t
  dt_1\int_0^{t} dt_2 \sum_{m,n} G^{R}_{im}(t-t_1)\Sigma_K^B(m,n,t_1-t_2)G^{A}_{n j}(t_2-t) .
  \eqa
\end{widetext}
For the sake of concreteness, we will now focus on a specific model of open quantum system with a microscopic form of $H_b$ and $H_{sb}$. For most of the calculation in the main text, unless it is mentioned otherwise, we have considered a widely used non-Markovian OQS model where each site of the system is coupled linearly to the first site of a semi-infinite bath of non-interacting Bosons hopping along the $1D$ chain with the tunneling strength $t_B$.

\begin{figure*}[t]
  \centering
   \includegraphics[width=0.24\textwidth , height =0.225\textwidth ]{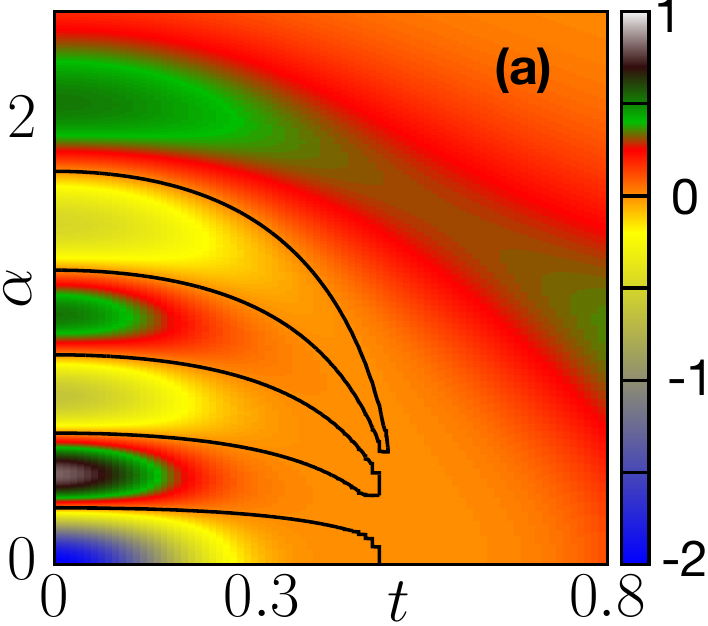} ~~
  \includegraphics[width=0.23\textwidth]{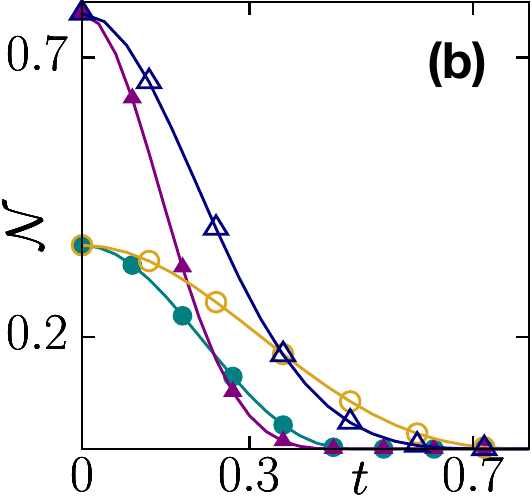} ~~
 \caption{(a)- (b) Evolution of a single mode non-interacting system coupled to a bath ($T=1$, $\mu=-4.05$)  starting from a Fock state: (a) The WQD, $W(\alpha,t)$ is shown in the $|\alpha|-t$ plane as a density plot. The system starts from the state $|n_i=5\rangle$. The negative patches present at the initial time, indicating a non-classical state, shrinks and vanishes as the system thermalizes to a classical state. (b) Negativity, $\mathcal{N}$ of WQD is plotted with $t$ for two different initial conditions $|n_i=6\rangle$ (triangles) and $|n_i=2\rangle$ (circles). In each case the open symbols correspond to a system bath coupling $\epsilon=1$, and the closed symbols are for $\epsilon=1.5$. Increasing $\epsilon$ makes the decay of $\mathcal{N}$ faster in both cases. For same $\epsilon$, the negativity for $n_i=6$ starts at higher value but decays faster than that for $n_i=2$.  $t_B=2$ sets the units for all plots. }
  \label{fig:onemode}
  \end{figure*}

We first consider a single mode harmonic oscillator \cite{Zhang1,Zhang2,Zhang4}, governed by $H_s=\omega_0 a^{\dagger}a$.  The system-bath coupling strength is governed by the energy scale $\epsilon$. We assume the system to start from a Fock state $|n_i \rangle \langle n_i|$. In this model, the bath spectral function is known to be of the form \cite{Chakraborty1},
\beq
 \mathcal{J}(\omega) = \Theta(4t_B^2-\omega^2)\frac{2}{t_{B}} \sqrt{1-\frac{\omega^2}{4t^2_{B}}},
\label{AP:Jomega_ourbath}
\eeq
which shows a square root derivative singularity at the two band edges $\omega=\pm 2t_B$. This non-analyticity in $\mathcal{J}(\omega)$ renders the reduced dynamics of the system to be non-Markovian which has been explored in great details in the reference \onlinecite{Chakraborty1}. This form of $\mathcal{J}(\omega)$ provide analytical solution for $\Sigma^R(\omega),\Sigma^K(\omega)$ and hence for $G^{R}(\omega)$, which are given by,
\bqa
\Sigma_R^B(\omega) &=& -\frac{\epsilon^{2}}{|t_{B}|}\left[  -\frac{\omega}{2~|t_{B}|} \pm\mathbf{i}~\sqrt{1- \frac{(\omega + \mathbf{i}\eta)^{2}}{4t_{B}^{2}}}      \right]  ,\nonumber \\
\Sigma_K^B(\omega) &=&  -\mathbf{i}\epsilon^2 J(\omega)~\coth \left( \frac{ \omega-\mu}{2T}\right) ,   \nonumber \\
G^R(t-t') &=& \Theta(t-t') \int \limits_{-\infty}^\infty \frac{d\omega}{2\pi} \frac{(-\mathbf{i} \epsilon^2)\mathcal{J}(\omega) e^{-\mathbf{i}\omega (t-t')}}{| \omega -\omega_0 -\Sigma_R^B(\omega)|^2}.
\eqa
In this case $\Lambda^0(t)=|G^R(t)|^2+\mathbf{i}\int_0^t dt_1\int_0^t dt_2 G^R(t-t_1)\Sigma^K(t_1-t_2)G^{R\ast}(t-t_2)$ and $\Lambda(t)=|G^R(t)|^2$.
The WQD $W(|\alpha|,t)$, defined as $W(\{\alpha_j\},t)= \int \prod_j \frac{d^2\beta_j}{\pi^N} \chi_W(\{\beta_j\},t) e^{\sum_j\beta_j^\ast\alpha_j
  -\beta_j \alpha_j^\ast}$ obtains the form,
\begin{equation}
\label{W_1mode}
W(\alpha,t) =\frac{2(\tilde{\Lambda}-\Lambda)^{n_i}}{(\Lambda^0)^{n_i+1}}e^{-\frac{2|\alpha|^2}{\Lambda^0}}L_{n_i}\left[\frac{2|\alpha|^2}{\Lambda^0} \frac{2\Lambda}{\Lambda-\tilde{\Lambda}}\right],
\end{equation}
where $\tilde{\Lambda}(t)=\Lambda^0(t)-\Lambda(t)$. $W(\alpha,0)=2e^{-2|\alpha^2|}L_{n_i}(4|\alpha^2|)$ has $n$ zero crossings in $|\alpha|$ resulting in alternating positive and negative patches. Here, $\Lambda(t)$ controls the decay of the initial distribution function, whereas $\tilde{\Lambda}(t)$ is the stochastic contribution from the bath. Since $L_{n_i}(x)>0$ for $x<0$, all the negative patches in $W$ disappear simultaneously 
when $\tilde{\Lambda}(t)>\Lambda(t)$ turns negative, i.e. when the stochastic contribution from the bath overwhelms the remnants of the initial quantum state.
We consider a system with $\omega_0=-1$ and a bath with $T=1$ and $\mu=-4.05$ where bath hopping scale,  $t_B=2$ sets the units. Fig.~\ref{fig:onemode}(a) shows a density plot of $W(\alpha,t)$ in the $|\alpha|-t$ plane for a system starting in the $n_i=5$ state. The $3$ negative patches shrink in size with time until they simultaneously disappear around $t=0.45$. 
In the long time limit, the system thermalizes and a gaussian WQD is recovered. A quantitative measure of the ``non-classicality'', the negativity, $\mathcal{N}(t) = \int d^2\alpha \left( |W(\alpha,t)|-W(\alpha,t) \right)/(4\pi)$, which measures the area in the complex $\alpha$ plane over which the WQD is negative. In Fig~\ref{fig:onemode}(b)  $\mathcal{N}$ is plotted as a function of $t$ for two different initial conditions,  $n_i=2$ and $n_i=6$,  at two different system-bath coupling strength $\epsilon=1, 1.5 $. The system has more non-classicality for higher $n_i$ as the number of negative patches increase, but the negativity also decays faster in this case. With increasing $\epsilon$, the negativity falls off faster.

\begin{figure*}[t]
  \centering
  \includegraphics[width=0.23\textwidth]{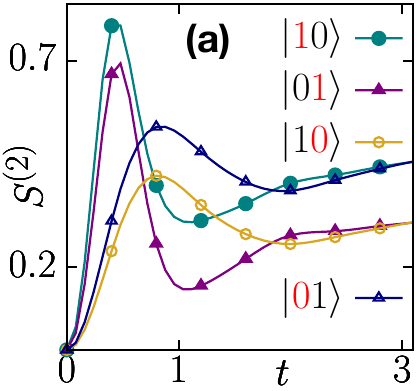} ~~~
  \includegraphics[width=0.23\textwidth]{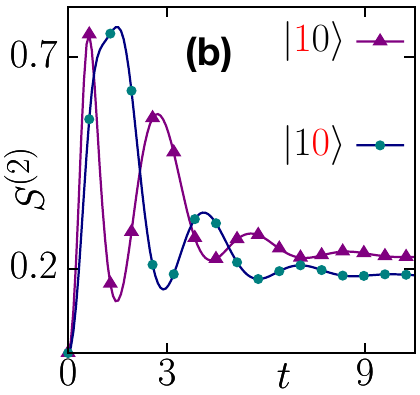} ~~~
   \includegraphics[width=0.24\textwidth,height=0.22\textwidth]{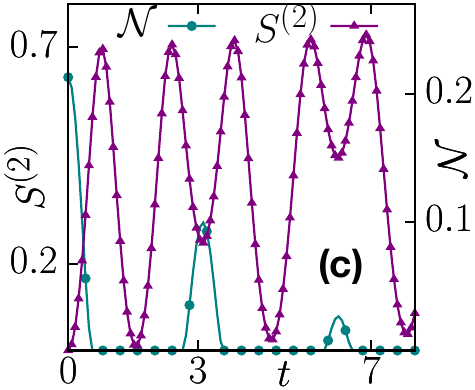}
 \caption{(a) $S^{(2)}$ vs $t$ in a $2-$mode system coupled to two baths ($T_L=T_R=1$, $\mu_L =-4.05, \mu_R=-10.05,\omega_0 =-1.5$) for $\epsilon=1.5>g=1$ (overdamped) and  (b)  $\epsilon=0.8<g=1.0$ (under-damped) for different initial states, marked in the graph.  The bath and the site index shown in black is traced out to obtain $S^{(2)}$. (c) $\mathcal{N}$ and $S^{(2)}$ for a system with $\epsilon=0.2$. The negativity shows a peak at alternate dips of $S^{(2)}$, where the density matrix comes close to $n_i=1$ Fock state. $t_B=2$ sets the units for all plots. }
  \label{fig:onemode}
  \end{figure*}

While the single mode system illustrates the effect of incoherent coupling to the bath, a more interesting dynamics occurs in a two mode system \cite{Zhang3}, where the coherent intermode coupling competes with the incoherent coupling to the bath. We next consider a system of two sites (left (L) and right (R)) with the Hamiltonian $H_s= \omega_0(a^\dagger_La_L+a^\dagger_Ra_R) -g (a^\dagger_La_R +h.c)$, where $g$ is the coherent tunneling matrix element. The two sites are linearly coupled to two baths with the same temperature $T$, but different chemical potentials $\mu_L$ and $\mu_R$. This leads to a steady state with a finite current flowing through the system in the long time limit. We study the evolution of $S^{(2)}$ in the dynamics of the OQS when the system starts from $|10\rangle$ and $|01\rangle$. The subsystem for calculating $S^{(2)}$ is obtained by tracing out the bath and one site of the system. We see that at short times, the behaviour of $S^{(2)}$ depends on whether the site with the initial particle is traced out or not. For example, in Fig~\ref{fig:onemode} (a), the curves for the system starting from $|10\rangle$ with $R$ traced out (closed circles) and $|01\rangle$ with $L$ traced out (closed triangles) follow each other at short times. Similarly, curves for system starting at $|10\rangle$ with $L$ traced out (open circles) and $|01\rangle$ with $R$ traced out (open triangles) follow each other at short times. At large times, this initial memory is erased and $S^{(2)}$ simply depends on whether $L$ or $R$ is traced over. This is because the chemical potentials in the baths break the $L-R$ symmetry of the problem.
Fig.~\ref{fig:onemode}(c) plots the evolution of $S^{(2)}$ and the negativity of the Wigner function of the reduced $\hat{\rho}$ for the left site in the under-damped limit. The system comes close to a Fock state at the minima of the entropy oscillations. When the corresponding state is $|0\rangle$, the negativity is zero, since the ground state has positive definite Wigner function, while the negativity shows a pronounced peak at alternate $S^{(2)}$ minima, when the system comes very close to $|1\rangle$.

\section{Effect of inter-particle interaction in a single mode OQS:}
\begin{figure}[t]
\includegraphics[width=1.0\textwidth]{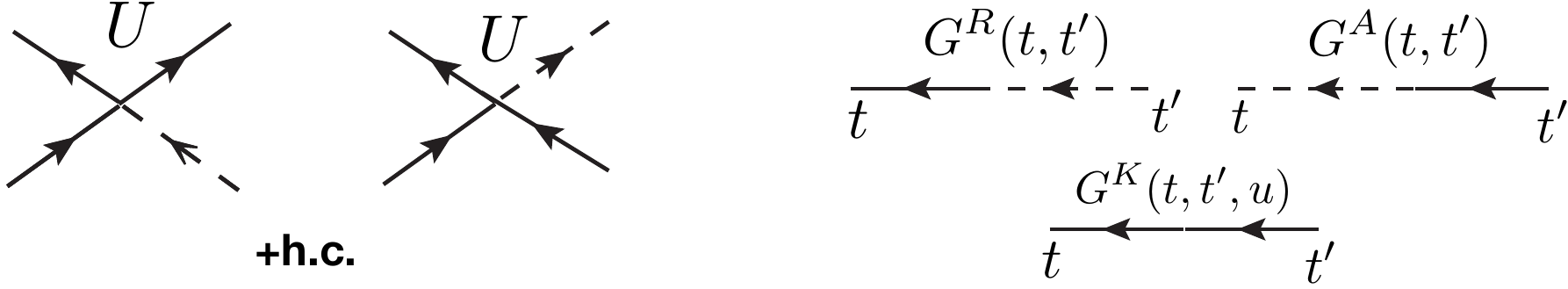}
\caption{Feynman interaction vertices and Green's functions.} 
\label{AP:vert}
\end{figure}

\begin{widetext}
In this section, we will extend the above formalism and detail the structure of the calculations in presence of an interparticle interaction (also called Kerr non-linearity)  in the dynamics of one mode open quantum system, given by
\beq
H_{int} = U a^\dagger a^\dagger a a.
\eeq
This leads to an interaction term in the reduced action of the system,

\beq
\mathcal{A}_{int} = -U \int dt \Big[\phi^{*}_{cl}(t)\phi_{cl}(t) \Big( \phi^{*}_{cl}(t) \phi_{q}(t) +  h.c \Big) +
\phi^{*}_{q}(t)\phi_{q}(t) \Big( \phi^{*}_{q}(t) \phi_{cl}(t) +  h.c \Big) \Big] .
\eeq
Now we will extend the definition of one particle Green's function $G^K(t,t,u)$ to introduce the connected $m$-particle Green's function of the classical fields at equal time $t$ as,
\beq
\Lambda^{(m)}(t,u) = \mathbf{i}^m G^{K(m)}(t,u) = (-1)^m \left. \frac{\partial^{2m}
  \log~Z[J_{cl},J_q,u]}{\partial{J_q(t)}..\partial{J_q(t)}\partial{J_q^\ast(t)}...\partial{J_q^\ast(t)}}\right. \Bigg\vert_{J_{cl/q}=0} .
\eeq
\end{widetext}
\begin{figure}[t]
\includegraphics[width=1.0\textwidth]{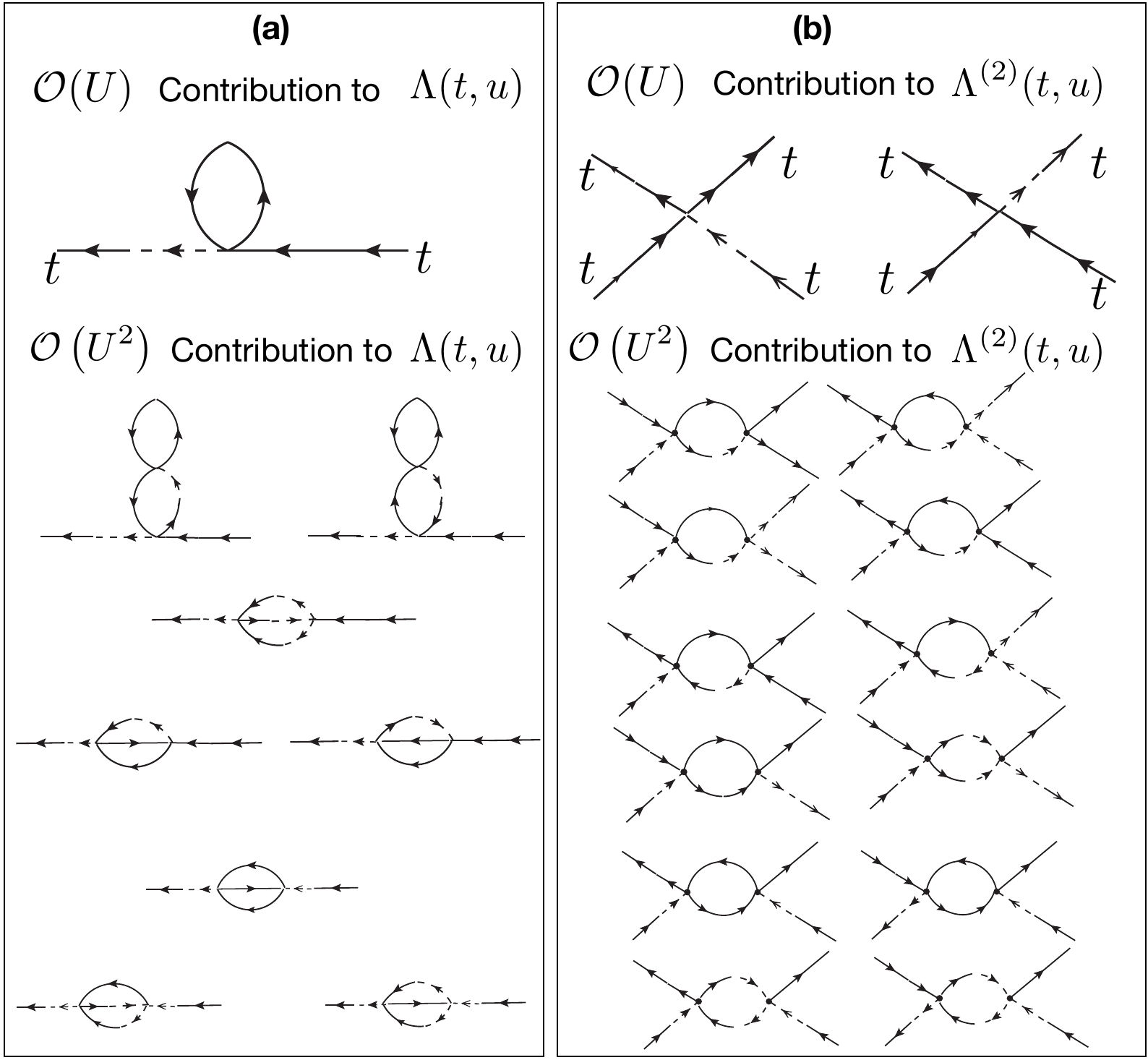}
\caption{Feynman Diagrams for $\Lambda(t,u)$ and $\Lambda^{(2)}(t,u)$.} 
\label{fig:lam2loop}
\end{figure}
Here, we would like to stress on a crucial point that for a system starting from non-thermal initial state, such as a Fock state, the $2m$ point connected Green's function, $\Lambda^{(m)}(t,u)$, with the initial source $u$ defined above, is not equal to the $m$ particle connected classical physical correlation function, defined by $\langle \phi_{cl}(t) .. \phi_{cl}(t) \phi^\ast_{cl}(t) .. \phi^\ast_{cl}(t)\rangle_C$. Rather they are related by appropriate derivative w.r.t $u$, dictated by $\hat{\rho}(0)$, as $\langle \phi_{cl}(t) .. \phi_{cl}(t) \phi^\ast_{cl}(t) .. \phi^\ast_{cl}(t)\rangle_C = \frac{1}{n!} \left(\frac{\partial}{\partial u} \right)^{n}\Lambda^{(m)}(t,u) $ (for details see \cite{chakraborty2}).
Using the fact that logarithm of partition function, $Z[J_{cl},J_q,u]$ is the generator of the connected correlator and the relation between WCF and the partition function (equation \ref{rel_chiWZ}), we write the WCF in presence of $u$ for an interacting system as,
\beq
\chi_{W}(\beta_j,t,u) = e^{\sum_{m=1}^{\infty} \frac{(-1)^m}{2^m~m!m!}  \Lambda^{(m)}(t,u)~ |\beta|^{2m}}.
\eeq
We will approximate the series in the exponent by truncating it at the order $m=2$, i.e. keeping upto $2-$ particle connected Green's function in the exponent. This approximation leads to analytical solution for Renyi entanglement entropy as follows,
\bqa
\label{AP:s2}
S^{(2)}&=&-\log \left[ 2\int_0^{\infty} d\beta~\beta~e^{-\Lambda^{(2)}(t,0)|\beta|^2 + \frac{1}{4}\Lambda^{(4)}(t,0)|\beta|^4}  \right]~\mathcal{M}\left( |\beta|^2\right) \nonumber \\
&=& -\log \left[ \sum \limits_p a_p \frac{2^{\frac{p+1}{2}}\Gamma(p+1)}{[-\Lambda^{(2)}(t,0)]^{\frac{p+1}{2}} } ~\mathcal{H}_{p+1}\left( \frac{\Lambda(t,0)} {\sqrt{-2\Lambda^{(2)}(t,0)}}\right)\right],
\eqa
where $\mathcal{M}\left( |\beta|^2\right)=\sum_p a_p |\beta|^{2p}$ is a polynomial in $|\beta|^2$, which is brought down by the $n^{th}$ order derivative w.r.t. $u$ in $\chi_W(\beta,t)$ given by equation \ref{AP:chi_W_deriv}. A closed form answer for $\mathcal{M}\left( |\beta|^2\right)$ can be obtained by using Bruno's identity\cite{grad}. In the regime of $U$ used in our calculation, we obtain $\Lambda^{(2)}(t,0) <0$, which regulates the integral in equation \ref{AP:s2}.

Now, we will sketch the structure of the perturbation theory to calculate $\Lambda(t,u)$ and $\Lambda^{(2)}(t,u)$ using standard diagrammatic expansion. The interaction vertices and the three kinds of Green's functions are shown in Fig. \ref{AP:vert}, while the Feynman diagrams for $\Lambda(t,u)$ and $\Lambda^{(2)}(t,u)$ upto $\mathcal{O}(U^2)$ are shown in Fig. \ref{fig:lam2loop}(a) and (b) respectively. In the diagrams, all the lines represent non-interacting Green's functions for the OQS in presence of the initial source, i.e. $G^{R/A}(t-t')$  and $G^K(t,t',u)$ given in equations \ref{AP:GR_single} and \ref{AP:Greens_GK_B}. In addition to the $2$nd order diagrams for $\Lambda^{(2)}$ shown in Fig. \ref{fig:lam2loop}(b), there are additional diagrams which can be obtained from the first order diagrams for $\Lambda^{(2)}$ by adding first order corrections to any one of the Green's functions.

As example of evaluating the expressions for the diagrams, we will write down here the explicit formula of $\Lambda(t,u)$,
\bqa
&&\Lambda(t,u) =-4 U ~Re[\int dt_1 ~G^R(t,t_1)  G^{K}(t_1,t_1,u)  G^K(t_1,t,u)]+ \nonumber \\
 &&2U^2 Im  \Big[ \int \limits_0^t dt_1  ~\int\limits_0^{t_1} dt_2 \bigg(4~   G^R(t,t_1)   G^K(t_1,t_2,u)  G^K(t_2,t_1,u)  G^R(t_1,t_2)  G^K(t_2,t,u) + \nonumber \\
&&2~   G^R(t,t_1)   G^K(t_1,t_2,u)  G^K(t_1,t_2,u)  G^A(t_2,t_1)  G^K(t_2,t,u) + 
2~   G^R(t,t_1)   G^R(t_1,t_2)  G^R(t_1,t_2)  G^A(t_2,t_1)  G^K(t_2,t,u) + \nonumber \\ 
&&4~   G^R(t,t_1)   G^K(t_2,t_2,u)  G^K(t_1,t_2,u)  G^A(t_2,t_1,u)  G^K(t_1,t,u) + 
4~   G^R(t,t_1)   G^K(t_2,t_2,u)  G^K(t_2,t_1,u)  G^R(t_1,t_2)  G^K(t_1,t,u) \bigg)
   \nonumber \\
&&+ \int \limits_0^t dt_1  ~\int\limits_0^{t} dt_2  G^R(t,t_1)   G^K(t_1,t_2,u)  G^K(t_2,t_1,u)  G^K(t_1,t_2,u)  G^A(t_2,t) +\nonumber \\
&& \int \limits_0^t dt_1  ~\int\limits_0^{t_1} dt_2 \left[2~ G^R(t,t_1)     G^K(t_2,t_1,u) G^R(t_1,t_2) G^R(t_1,t_2)  G^A(t_2,t) +
 4~ G^R(t,t_1)     G^K(t_1,t_2,u) G^R(t_1,t_2) G^A(t_2,t_1)  G^A(t_2,t) \right]  + \nonumber \\
&&    \int \limits_0^t dt_1  \int\limits_0^{t_1} dt_2 4  G^R(t,t_1)   G^K(t_1,t_1,u)  G^K(t_2,t_2,u)  G^R(t_1,t_2)  G^K(t_2,t,u) +\nonumber \\
  && \int \limits_0^t dt_1  \int\limits_0^{t} dt_2 2  G^R(t,t_1)   G^K(t_1,t_1,u)  G^K(t_2,t_2,u)  G^K(t_1,t_2,u)  G^A(t_2,t) \Big].
\eqa
\begin{figure}[t]
\centering
   \includegraphics[width=0.3\textwidth]{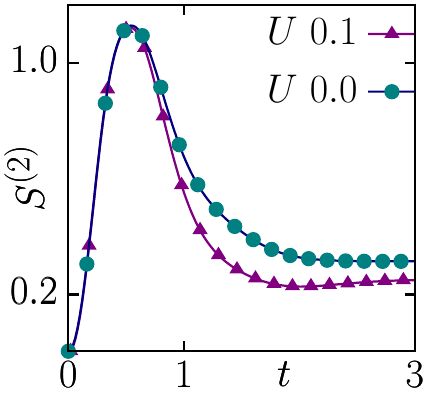}
\caption{$S^{(2)}$ vs $t$ in presence of repulsive interaction, $U=0.1$ starting from $n=2$ with $\epsilon=1.5$. Repulsive interaction leads to faster decay of $S^{(2)}$ compared to the non-interacting one, leading to a steady state with lower entropy.} 
\label{fig:onemode_int}
\end{figure}
Similar expressions for $\Lambda^{(2)}$ can be obtained by evaluating the diagrams \ref{fig:lam2loop}(b). 
We calculate the time dependent Renyi entropy of this system after integrating out the bath. The time evolution of $S^{(2)}$ for interacting and non-interacting systems are plotted together in Fig.~\ref{fig:onemode_int}. The repulsive interaction reduces the long time value of the entanglement entropy and leads to a faster decay of $S^{(2)}$ from its peak value.

\section{Non-monotonic evolution of $S^{(2)}$ with time}
In the main the text, we have shown that the non-monotonicity in $S^{(2)}$ in case of a single mode OQS starting from a Fock state, is not a particular feature of the non-Markovian model of bath spectral function (equation \ref{AP:Jomega_ourbath}) we have considered. Rather, this phenomenon is robust to Markovian bath model as well which we have illustrated in the main text. 

In this appendix, we will provide a physical intuition behind the non-monotonic evolution of $S^{(2)}$ applicable to a generic OQS (for both Markovian and non-Markovian dynamics). We propose an ansatz to write the time evolution of the density matrix which has two contributions: one coming from the decay of the initial state due to the bath induced dissipation and the other coming from an effective backward (advanced) propagation in time from the steady state density matrix $\hat{\rho}(\infty)$. This gives,
\begin{equation}
\hat{\rho}(t)=A(t) |n_i\rangle \langle n_i | + B(t) \hat{\rho}(\infty).
\label{rho_ansatz_generic}
\end{equation}
Here the coefficient of the retarded propagation $A(t)$ decays from $1$ at $t=0$ to $0$ at long time and $B(t)$ is the coefficient of the effective advanced propagation. It is clear that in the limit $t \rightarrow 0$, $B(t)\rightarrow 0$ and in the long time limit $t \rightarrow \infty$, $B(t) \rightarrow 1$. Within this ansatz the Renyi entropy evolves as, 
\beq
e^{-S^{(2)}(t)}=A^2(t) + B^2(t) e^{-S^{(2)}(\infty)}+ 2 A(t) B(t)  \rho_{n_i,n_i}(\infty).
\label{S_ansatz}
\eeq
Here, $ \rho_{n_i,n_i}(\infty)$, coming from the cross term in Eq.\ref{rho_ansatz_generic}, is the matrix element of the steady state thermal density matrix between the initial Fock state. 

We will now analyze the short-time behaviour of $S^{(2)}$ within this ansatz. At the short-time, we can heuristically assume (motivated by Fermi-Golden rule) a linearly decaying form of $A(t) \sim 1- at$ and also a linear growth in $B(t) \sim b t$ at rate $b$ whose exact form is hard to determine for a generic OQS. Substituting these forms in Eq.\ref{S_ansatz}, we get,
\begin{equation}
e^{-S^{(2)}(t)}=1-2a t \left(1-\rho_{n_i,n_i}(\infty) \frac{b}{a} \right).
\end{equation}
This immediately implies that $\rho_{n_i,n_i}(\infty)$, which contains information about both the initial state and the steady state, crucially governs the initial entropy production rate. This initial rise in $S^{(2)}(t)$ continues upto a timescale $t_{0}\sim 1/(2a)$ at which the memory of the initial conditions decay due to the dissipation coming from the bath. Thus at $t=t_0$, the Renyi entropy, within the linear approximation, is given by,
\begin{equation}
e^{-S^{(2)}(t=\frac{1}{2a})}= \rho_{n_i,n_i}(\infty) \frac{b}{a}.
\end{equation}
The evolution of  $S^{(2)}(t)$ will be non-monotonic if $S^{(2)}(t_0) > S^{(2)}(\infty)$ where the latter can be written in terms of the final steady state average density $n_f$ as, $exp \{-S^{(2)}(\infty) \}=1/(1+2n_f)$. This yields a criterion for a peak to appear in the evolution of $S^{(2)}(t)$ as,
\beq
   \frac{b}{a}\rho_{n_i,n_i}(\infty)< e^{-S^{(2)}(\infty)}=\frac{1}{1+2n_f}.
 \label{peak_generic}
\eeq
 Keeping $n_i$ fixed, if we increase $n_f$, the steady state thermal distribution broadens leading to an increase in $S^{(2)}(\infty)$. Hence, heuristically increasing $n_f$ leads to disappearance of the peak in $S^{(2)}(t)$.

We will now apply the general physical intuition behind the peak in $S^{(2)}(t)$ to the concrete example of the Markovian dynamics of the single mode OQS. In this case, we have not derived the Green's functions from a microscopic form of $\mathcal{J}(\omega) $. Instead, we have considered the generic form of the time-local dissipation and noise kernel under Markovian dynamics given by,
\bqa
\Sigma_R^B(t-t') = \mathbf{i} \gamma \delta(t-t') ~,~
\Sigma_K^B(t-t') =- \mathbf{i} D \delta(t-t') ,\nonumber
\eqa
which leads to exponentially decaying Green's functions,
\bqa
\label{AP:GR_single}
&&G^R(t-t')= -\mathbf{i}\Theta(t-t') e^{-\gamma (t-t')-\mathbf{i}\omega_0(t-t')} \nonumber \\
&&\Lambda^0(t)=e^{-2\gamma t} +\frac{ D}{2\gamma} [1-e^{-2\gamma t}],
\eqa
where $D/2\gamma=1+2n_f$ is related to the steady state number in the system and hence dictated by the temperature and the chemical potential of the bath. We have used $\gamma= 0.225$ for all the plots in the main text.

To analyze the peak in time-evolving $S^{(2)}(t)$ for a single mode OQS coupled to Markovian bath we propose an ansatz for the instantaneous reduced density matrix of the single-mode system, $\hat{\rho}(t)$,
\beq
\hat{\rho}(t)= e^{-2\gamma t} |n_i\rangle \langle n_i | + (1-e^{-2\gamma t}) \hat{\rho}(\infty),
\label{rho_ansatz_M}
\eeq
where the first term denotes the decay of the initial state governed the bath induced dissipation scale $\gamma$ and in the last term $\hat{\rho}(\infty)$ is the steady-state density matrix of the reduced single-mode system. Within this ansatz,
\beq
e^{-S^{(2)}(t)}=e^{-4\gamma t} + (1-e^{-2\gamma t})^2 e^{-S^{(2)}(\infty)} + 2 e^{-2\gamma t}  (1-e^{-2\gamma t})  \rho_{n_i,n_i}(\infty).
\eeq

Now, $S^{(2)}(t)$ shows a peak during its evolution in time when $dS^{(2)}/dt=0$ at a time $t$ given by $exp(-2\gamma t)= [ e^{-S^{(2)}(\infty)}-    \rho_{n_i,n_i}(\infty)]/[1+e^{-S^{(2)}(\infty)}-2   \rho_{n_i,n_i}(\infty)]$. This yields a criterion for the peak to occurs at a finite time $0 <t < \infty$ when,
\beq
  \rho_{n_i,n_i}(\infty)< e^{-S^{(2)}(\infty)}.
 \label{peak}
\eeq
At this point, we would like to note that the general criterion given in Eq.\ref{peak_generic} reduces to the above expression for the simplest Markovian dynamics if we set $b=a$.

To obtain a physical understanding of $ \rho_{n_i,n_i}(\infty)$ we will now calculate the matrix elements $\rho_{nn}(t)$ of the instantaneous reduced density matrix of the system $\hat{\rho}(t)$ in the number basis .
For a single mode reduced density matrix, $\chi_W(\beta,t) = \chi_W(|\beta|,t)$ is only a function of $|\beta|$ and $\hat{\rho}(t)$ is diagonal in number basis.
We constructed the matrix elements, $\rho_{nn}(t)$ from WCF through the following equation \cite{GlauberCahill},
\begin{eqnarray}
\chi_W(|\beta|,t)&=&\sum_{n} \rho_{nn} (t)~\langle n | e^{\beta~a^{\dagger} - \beta^* ~a} | n \rangle , \nonumber \\
\rho_{nn}(t) &=& \frac{1}{\pi}~\int d^2\beta~\chi_W(|\beta|,t)~\langle n | e^{\beta~a^{\dagger} - \beta^* ~a} | n \rangle,
\end{eqnarray}
where $\langle n | e^{\beta~a^{\dagger} - \beta^* ~a} | n \rangle$ is obtained to be,
\begin{equation}
\langle n | e^{\beta~a^{\dagger} - \beta^* ~a} | n \rangle=e^{- \frac{1}{2}|\beta|^2} ~ L_n[|\beta|^2].
\end{equation}
This leads to the closed form analytic answer for the matrix elements, $\rho_{nn}(t)$, for a system starting from a Fock state $|n_i \rangle \langle n_i|$, given as,
\begin{equation}
\rho_{nn}(t) = {}^{n_i+n}C_{n} \frac{2\left( \Lambda_0 + 1 -2 \Lambda \right)^{n_i}~\left( \Lambda_0 - 1 \right)^n}{\left( \Lambda_0 + 1 \right)^{n_i+n+1}} ~{}_{2}F_1~\left[-n,-n_i,-n_i-n,  \frac{\left( \Lambda_0 + 1 \right) \left( \Lambda_0 - 1 - 2\Lambda  \right) }{\left( \Lambda_0 - 1 \right)\left( \Lambda_0 + 1 - 2\Lambda  \right)} \right].
\end{equation}

Now, in the limit $t \rightarrow \infty$, $\Lambda  \rightarrow 0$ and $\Lambda_0 \rightarrow (1+2n_f)$ we get $ \rho_{n_i,n_i}(\infty) = n_f^{n_i}/(1+n_f)^{n_i+1}$. Hence the criterion for non-monotonic evolution of $S^{(2)}(t)$ given in Eq. \ref{peak} reduces to,
\beq
r=\frac{n_f^{n_i}}{(1+n_f)^{n_i+1}} (1+2n_f)<1,
\eeq
while for a generic OQS (given in Eq.\ref{peak_generic}) $b r /a <1$.

\bibliographystyle{apsrev4-1}
\bibliography{Wignerfn.bib}